\documentclass[aps,prb,twocolumn,superscriptaddress,showpacs,floatfix]{revtex4}
\usepackage{graphicx,color}
\usepackage{dcolumn}
\usepackage{hyperref}
\usepackage{amsmath}
\hfuzz=\maxdimen
\usepackage{float}
\begin{document}

\title{Strong Pauli paramagnetic effect in the upper critical field of KCa$_2$Fe$_4$As$_4$F$_2$}


\author{Teng Wang}
\affiliation{State Key Laboratory of Functional Materials for
Informatics, Shanghai Institute of Microsystem and Information
Technology, Chinese Academy of Sciences, Shanghai 200050,
China}\affiliation{CAS Center for Excellence in Superconducting
Electronics(CENSE), Shanghai 200050, China}\affiliation{School of Physical Science and Technology, ShanghaiTech University, Shanghai 201210, China}

\author{Chi Zhang}
\affiliation{State Key Laboratory of Functional Materials for
Informatics, Shanghai Institute of Microsystem and Information
Technology, Chinese Academy of Sciences, Shanghai 200050,
China}\affiliation{CAS Center for Excellence in Superconducting
Electronics(CENSE), Shanghai 200050, China}\affiliation{University
of Chinese Academy of Science, Beijing 100049, China}

\author{Liangcai Xu}\affiliation{Wuhan National High Magnetic Field Center and School of Physics,
Huazhong University of Science and Technology, Wuhan 430074, China}

\author{Jinhua Wang}\affiliation{Wuhan National High Magnetic Field Center and School of Physics,
Huazhong University of Science and Technology, Wuhan 430074, China}

\author{Shan Jiang}\affiliation{Wuhan National High Magnetic Field Center and School of Physics,
Huazhong University of Science and Technology, Wuhan 430074, China}

\author{Zengwei Zhu}\affiliation{Wuhan National High Magnetic Field Center and School of Physics,
Huazhong University of Science and Technology, Wuhan 430074, China}

\author{Zhaosheng Wang}\affiliation{Anhui Province Key Laboratory of Condensed Matter Physics at Extreme Conditions, High Magnetic Field Laboratory of the Chinese Academy of Sciences, Hefei 230031, China}

\author{Jianan Chu}
\affiliation{State Key Laboratory of Functional Materials for
Informatics, Shanghai Institute of Microsystem and Information
Technology, Chinese Academy of Sciences, Shanghai 200050,
China}\affiliation{CAS Center for Excellence in Superconducting
Electronics(CENSE), Shanghai 200050, China}\affiliation{University
of Chinese Academy of Science, Beijing 100049, China}

\author{Jiaxin Feng}
\affiliation{State Key Laboratory of Functional Materials for
Informatics, Shanghai Institute of Microsystem and Information
Technology, Chinese Academy of Sciences, Shanghai 200050,
China}\affiliation{CAS Center for Excellence in Superconducting
Electronics(CENSE), Shanghai 200050, China}\affiliation{University
of Chinese Academy of Science, Beijing 100049, China}

\author{Lingling Wang}
\affiliation{State Key Laboratory of Functional Materials for
Informatics, Shanghai Institute of Microsystem and Information
Technology, Chinese Academy of Sciences, Shanghai 200050, China}

\author{Wei Li}
\affiliation{Department of Physics and State Key Laboratory of Surface Physics, Fudan University, Shanghai 200433, China}

\author{Tao Hu}
\affiliation{State Key Laboratory of Functional Materials for
Informatics, Shanghai Institute of Microsystem and Information
Technology, Chinese Academy of Sciences, Shanghai 200050,
China}\affiliation{CAS Center for Excellence in Superconducting
Electronics(CENSE), Shanghai 200050, China}

\author{Xiaosong Liu}
\affiliation{State Key Laboratory of Functional Materials for
Informatics, Shanghai Institute of Microsystem and Information
Technology, Chinese Academy of Sciences, Shanghai 200050,
China}\affiliation{CAS Center for Excellence in Superconducting
Electronics(CENSE), Shanghai 200050, China}\affiliation{School of Physical Science and Technology, ShanghaiTech University, Shanghai 201210, China}

\author{Gang Mu}
\email[]{mugang@mail.sim.ac.cn} \affiliation{State Key Laboratory of
Functional Materials for Informatics, Shanghai Institute of
Microsystem and Information Technology, Chinese Academy of Sciences,
Shanghai 200050, China}\affiliation{CAS Center for Excellence in
Superconducting Electronics(CENSE), Shanghai 200050, China}





\begin{abstract}
Recently, 12442 system of Fe-based superconductors has attracted considerable attention owing to its unique double-FeAs-layer structure.
A steep increase in the in-plane upper critical field with cooling has been observed near the superconducting transition temperature, $T_c$, in KCa$_2$Fe$_4$As$_4$F$_2$ single crystals.
Herein, we report a high-field investigation on upper critical field of this material over a wide temperature range, and both out-of-plane ($H\|c$, $H_{c2}^{c}$) and in-plane ($H\|ab$, $H_{c2}^{ab}$) directions have been measured.
A sublinear temperature-dependent behavior is observed for the out-of-plane $H_{c2}^{c}$, whereas strong convex curvature with cooling is observed for the in-plane $H_{c2}^{ab}$.
Such behaviors could not be described by the conventional Werthamer--Helfand--Hohenberg (WHH) model. The data analysis based on the WHH model by considering the spin aspects reveals a large
Maki parameter $\alpha=9$, indicating that
the in-plane upper critical field is affected by a very strong Pauli paramagnetic effect.


\end{abstract}


\maketitle


\section*{1 Introduction}
Unconventional superconductivity is closely related to the electronic and crystal structures of the materials~\cite{Xu2018}. Adjustment of the crystal structure often leads to abundant physical manifestations~\cite{LaFeAsO,Zhu2009,FeSe,122}.
From this viewpoint, the 12442 system, which is the only system with two FeAs layers between neighboring insulating layers in Fe-based superconductors (FeSCs), has gained considerable attention
in the field of superconducting (SC) materials~\cite{12442-1,12442-2,12442-3,12442-4,12442-5,Wang2016,Ishida2017,Kirschner2018,Adroja2018,Wang2019}. Depending on the detailed compositions, materials in this system
are superconducting with $T_c$ =28--37 K~\cite{12442-1,12442-2,12442-3,12442-4,12442-5}.
The recent advances in the field of high-quality single crystal growth have greatly promoted studies on the intrinsic physical properties of this system~\cite{WangCrystal,Huang2019,TWang2019}. For example,
a large anisotropy of the upper critical field $\Gamma=H_{c2}^{ab}/H_{c2}^{c}$ and very steep increase of the in-plane $H_{c2}^{ab}$ with cooling are observed in the temperature region near $T_c$
in CsCa$_2$Fe$_4$As$_4$F$_2$ and KCa$_2$Fe$_4$As$_4$F$_2$ single crystals~\cite{WangCrystal,TWang2019}.
These findings may be significant both for the fundamental physics and potential applications.
Currently, however, the behavior at lower temperatures and higher magnetic fields is still unknown.

Generally, for type-II
superconductors, the upper critical field can be affected by the interactions between the external magnetic field and orbital motion of the SC electrons, by the effect of the
field on the electron spin magnetic moments (Pauli paramagnetic effect), and by the spin-orbit scattering~\cite{Maki1964,Maki1966,WHH-1,WHH-2}. In the case of FeSCs, the Pauli paramagnetic effect is significant,
severely limiting the in-plane $H_{c2}^{ab}$ when approaching zero temperature~\cite{Yuan2009,Fang2010,Khim2010,Khim2011,Xing2017,ZSWang2017}.
Therefore, almost all the FeSCs exhibit isotropic characteristics in low temperature. Hence, the behavior of the upper critical field of 12442 system has gained great research interest.
The pulsed magnetic field can provide a vital platform for studying this subject~\cite{YHMa2018}.
In this study, we conducted an in-depth examination of the upper critical field of KCa$_2$Fe$_4$As$_4$F$_2$ single crystals using a pulsed magnetic field of up to 50 T. Strongly convex curvature with cooling
was observed in the $H_{c2}^{ab}-T$ curve; this convexity interrupts the sharp upward trend of $H_{c2}$ near $T_c$. The Werthamer--Helfand--Hohenberg model considering the spin aspects fitted the experimental 
data, and a significantly strong Pauli paramagnetic effect was revealed.

\begin{figure*}
\includegraphics[width=15cm]{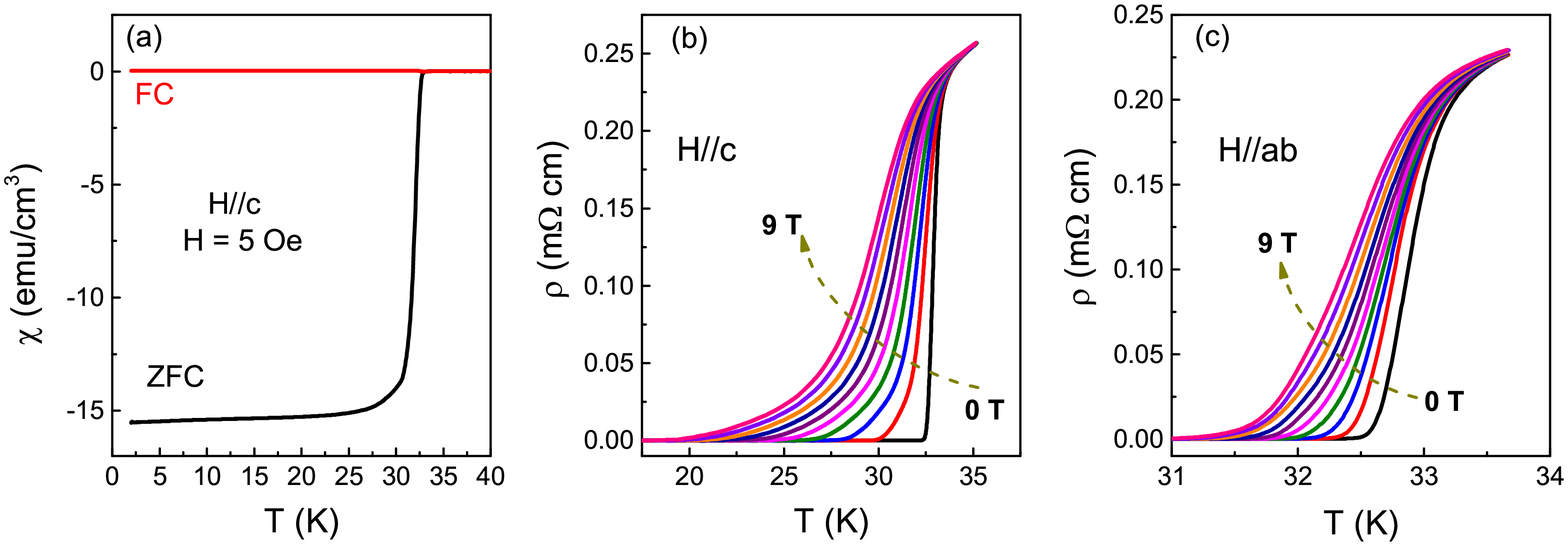}
\centering \caption {(a) Magnetic susceptibility measured in zero-field-cooled
(ZFC) and field-cooled (FC) modes. An external magnetic field of 5 Oe is applied along the $c$ axis. (b)--(c) The temperature dependence of electronic resistivity under different magnetic fields up to 9 T with $H\|c$ and
$H\|ab$, respectively. The interval for the field is 1 T/step.}
\label{fig1}
\end{figure*}

\section*{2 Materials and methods}\label{sec:4}
The KCa$_2$Fe$_4$As$_4$F$_2$ single crystals were grown using the KAs self-flux
method~\cite{TWang2019}. CaF$_2$ powder and
homemade CaAs, KAs, and Fe$_2$As were mixed together in a stoichiometric ratio and placed in an alumina crucible; then, they were sealed in a stainless steel pipe~\cite{Kihou2010}.
Fifteen times excess KAs was added as the flux. Moreover, appropriate amounts of CaAs and CaF$_2$ were added to suppress the formation of 122 phase KFe$_2$As$_2$. The materials were heated at
980 $^{\circ}$C for 20 h and cooled down to 900 $^{\circ}$C at a rate of
1.6--4 $^{\circ}$C/h, followed by rapid cooling to room
temperature. The detailed characterizations of the samples has been reported in our previous work~\cite{TWang2019}.

\begin{figure}[b]
\includegraphics[width=6.2cm]{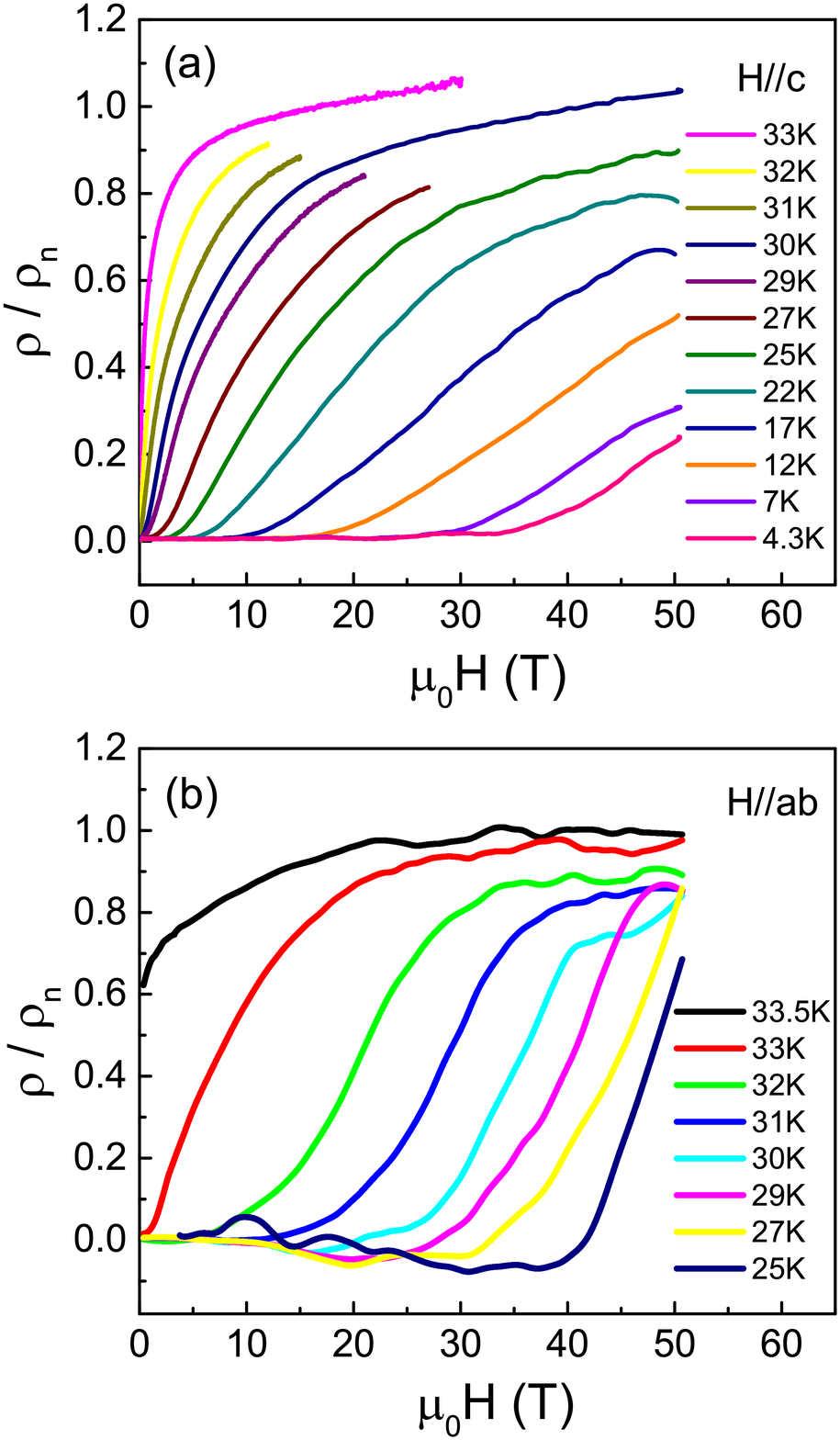}
\caption {Magnetic field dependence of resistivity at different
temperatures for KCa$_2$Fe$_4$As$_4$F$_2$. The data are normalized to the value in the normal state $\rho_n$. The magnetic field is applied with two orientations (a) $H\|c$ and (b) $H\|ab$.} \label{fig2}
\end{figure}

Electronic resistivity was measured with a DC magnetic field of up to
9 T on a physical property measurement system (PPMS, Quantum Design).
The experiments were also conducted under the pulsed magnetic fields of up to 50 T. The magnetic fields were applied along the $c$-axis ($H\|c$) and $ab$-plane ($H\|ab$) of the crystals.
Current was applied within the ab-plane and was perpendicular to the direction of the magnetic field. Magnetic susceptibility measurement was
conducted using a magnetic property measurement system (MPMS, Quantum Design).

\section*{3 Results and discussion}
In our previous work~\cite{TWang2019}, we reported that $\chi-T$ data measured with $H\|ab$ represents a superconducting volume fraction very close to 100\%.
Herein, we show the temperature dependence of magnetic susceptibility data $\chi-T$ with $H\|c$, see Fig. 1(a). The data reveals a very sharp SC transition at $T_c$ = 33.2 K,
indicating that our sample is of high quality and good homogeneity.
The in-plane resistivity as a function of temperature was measured under different fields up to 9 T. In Figs. 1(b) and (c), we show the $\rho-T$ data with $H\|c$ and $H\|ab$, respectively. The SC
transition displays a width of approximately 1 K under zero field, which is broadened by the external magnetic field. Such a broadening has been reported in 1111 system of FeSCs and cuprates, which is a consequence of the formation
of a vortex-liquid state~\cite{Jaroszynski2008,Lee2009,Kwok1992,Safar1992}. Note that the scale of
the $x$-coordinates for these two graphs are different, representing a clear anisotropy in the suppression of the SC transition by the magnetic field. Because of the limitation of the magnetic field, we only obtained
information in a very narrow temperature region near $T_c$. Hence, we conducted transport measurements under the pulsed magnetic fields of up to 50 T.

\begin{figure}
\includegraphics[width=8.5cm]{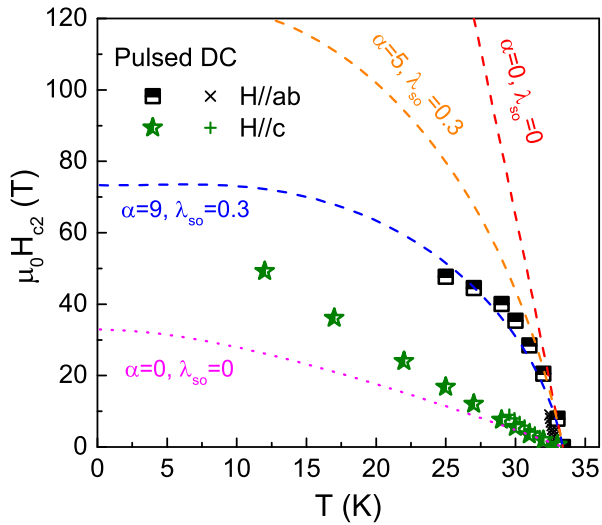}
\caption {Upper critical field $\mu_0H_{c2}(T)$ versus temperature for KCa$_2$Fe$_4$As$_4$F$_2$ single crystals. Half-filled symbols and crosses represent the data obtained
in the pulsed field and low magnetic field, respectively. The dashed and dotted lines are the WHH fits for the experimental data with $H\|ab$ and $H\|c$, respectively.} \label{fig3}
\end{figure}

Normalized resistivity as a function of pulsed magnetic field is shown in Figs. 2(a) and (b). With the decrease of temperature, the SC state can survive under a higher field. In the case of $H\|c$ (see Fig. 2(a)),
the SC transition becomes broadening in the low temperature region, indicating a wide area of the vortex-liquid state. This is consistent with the observations shown in Fig. 1(b).
As shown in Fig. 2(b), such a broadening behavior is not
observed in the case of $H\|ab$. This difference in behavior is a consequence of the different types of vortices in the different field orientations in a quasi-two-dimensional layered material~\cite{Blatter1994}.
It is believed that the presence of SC fluctuation near the
onset of the SC transition and vortex-liquid state near zero-resistivity will affect the determination of upper critical field~\cite{Xing2017,ZSWang2017}. Hence, the criteria 50\%$\rho_n$ was adopted herein.
The normal state resistivity $\rho_n$ was determined from the $\rho-T$ curve under zero field.

The upper critical fields for the two orientations, $H_{c2}^{ab}$ and $H_{c2}^c$, as a function of temperature are shown in Fig. 3. The data from the pulsed field and DC field experiments are roughly consistent.
We first checked the behavior of in-plane $H_{c2}^{ab}(T)$. Adjacent to the SC transition $T_c$, $H_{c2}^{ab}(T)$ shows a steep increase with cooling leading to a slope $H'=-d\mu_0H_{c2}^{ab}/dT$$\mid$$_{T_c}$ = 19.4 T/K.
This value is smaller than that obtained using the criteria 90\%$\rho_n$~\cite{TWang2019} because of the broadening effect induced by the magnetic field. As the temperature decreases further, this steep tendency is
interrupted severely, and the $H_{c2}^{ab}-T$ curve reveals strongly convex curvature in the temperature region 25--30 K.
On the contrary, the out-of-plane $H_{c2}^c(T)$ shows a linear temperature-dependent behavior in fields exceeding 8 T. Below 8 T, the $H_{c2}^{c}-T$ curve displays a tail-like feature with a low slope.
Similar features are also observed in other systems of FeSCs~\cite{Xing2017,ZSWang2017,Pisoni2016}.

To comprehensively evaluate the effects of the orbital pair-breaking effect, spin-paramagnetic pair-breaking effect, and spin-orbit interaction on the upper critical field, we checked our
data based on the WHH theory~\cite{WHH-2}. In the dirty limit, according to the WHH theory, the upper critical field can be obtained from the function~\cite{WHH-2}
\begin{equation}
\begin{split}
ln\frac{1}{t}=&\sum_{\nu=-\infty}^{\infty}\{\frac{1}{|2\nu+1|}-[|2\nu+1|\\&+\frac{\bar{h}}{t}+\frac{(\alpha\bar{h}/t)^2}{|2\nu+1|+(\bar{h}+\lambda_{so})/t}]^{-1}\}, \label{eq:1}
\end{split}
\end{equation}
where $t=T/T_c$, $\bar{h}=4H_{c2}/(\pi^2H'T_c)$, and $\alpha$ and $\lambda_{so}$ are parameters reflecting the strength of the spin paramagnetic effect and spin-orbit interaction, respectively. The parameter $\alpha$ was proposed by Maki,
known as Maki parameter~\cite{Maki1964}.
For the orientation of $H\|ab$, the conventional scenario without spin-paramagnetic effect and spin-orbit interaction ($\alpha$ = 0, $\lambda_{so}$ = 0) could not describe the experimental data (see the red dashed line in Fig. 3); the conventional scenario gives an estimation exceeding thrice the experimental value at 25 K.
By adjusting the values of $\alpha$ and $\lambda_{so}$, the data can be roughly represented by the theoretical model with the parameters $\alpha$ = 9 and $\lambda_{so}$ = 0.3, as shown by blue dashed line in Fig. 3.
Note that the value of $\alpha$ obtained herein is very large, and the previous maximum value of $\alpha$ = 6.5 is found in the 11111 system~\cite{ZSWang2017}. This indicates a very strong spin-paramagnetic effect in the present system.
Meanwhile, the spin-orbit interaction with a strength $\lambda_{so}$ = 0.3 is essential to describe the experimental data. To give a vivid impression for the necessity of such a large $\alpha$, we also show the theoretical
curve with $\alpha$ = 5 and $\lambda_{so}$ = 0.3, as shown by yellow dashed line in Fig. 3, which fails to reflect the strongly convex curvature at approximately 25 K.

In the case of out-of-plane case ($H\|c$), as shown by pink dotted line in Fig. 3, the conventional WHH model again fails to describe the experimental data. Moreover, the theoretical curve is below the experimental data. This means
that the introduction of $\alpha$ and $\lambda_{so}$ can not improve the situation because typically the spin-paramagnetic effect further suppresses the value of the upper critical field~\cite{WHH-2}.
We note that this sublinear feature of the $H_{c2}^c-T$ is common in the family of FeSCs and is widely believed to be a consequence of the multiband effect~\cite{Xing2017,ZSWang2017}.
The feature of multiband effect is more pronounced for the orientation with $H\|c$ than that with $H\|ab$. The ratio $\eta = D_h/D_e$, where $D_h$ and $D_e$ are the diffusivities of the hole and electronic bands
respectively, is the key factor controlling the $H_{c2}-T$ behavior of the multiband superconductors in the dirty limit~\cite{Gurevich2003,Gurevich2007,Hunte2008}. Typically, the in-plane value $D_h^{ab}$ is much smaller than $D_e^{ab}$
because of the large effective mass of the hole-type carriers. When the field is applied along the $c$ axis, only the in-plane diffusivity is involved. Hence, we obtain a quite small value of $\eta$,
leading to the pronounced feature of the multiband effect.
For $H\|ab$, on the contrary, the out-of-plane values of $D_h^{c}$ and $D_e^{c}$ should be considered when calculating $\eta$, and hence, $\eta$ is enhanced.
This may explain the observed indistinctive multiband feature in the case of $H\|ab$. Such a tendency was noticed and discussed by Hunte et al. when studying the LaFeAsO$_{0.89}$F$_{0.11}$ system~\cite{Hunte2008}.

\begin{figure}
\includegraphics[width=6.8cm]{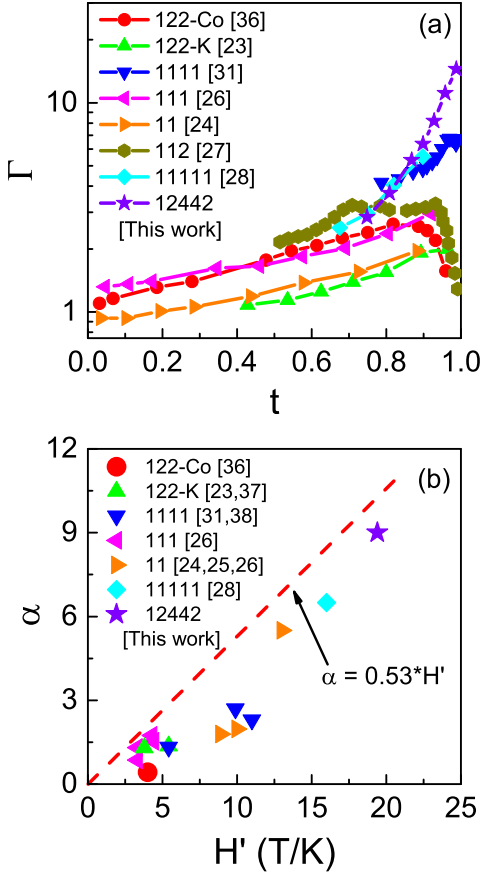}
\caption {(a) Anisotropic parameter $\Gamma=H_{c2}^{ab}/H_{c2}^c$ as a function of the reduced temperature $t=T/T_c$.
(b) The spin Maki parameter $\alpha$ as a function of the slope $H'=-d\mu_0H_{c2}^{ab}/dT$$\mid$$_{T_c}$. Red dashed line denotes the Maki relation in the weak-coupling limit. The references for the data are denoted in the
figure.} \label{fig4}
\end{figure}

Finally, we performed a quantitative comparison of the studied system with other systems of FeSCs~\cite{Yuan2009,Kano2009,Terashima2009,Lee2009,Fuchs2008,Khim2011,Fang2010,Khim2010,Xing2017,ZSWang2017}.
In Fig. 4(a), the anisotropic parameter $\Gamma$ is plotted as a function of the reduced temperature $t=T/T_c$ for different FeSCs including
122, 1111, 111, 11, 112, 11111, and the present 12442 system. It is interesting
that all the curves share a common tendency toward $\Gamma\simeq 1$ as the temperature approaches 0 K. The present KCa$_2$Fe$_4$As$_4$F$_2$
system also follows this general trend. The difference in the case of 12442 material is that it has a larger value of $\Gamma$ near $T_c$ ($t=1$), which reveals a steeper decrease with cooling.
Next, we checked the Maki parameter $\alpha$.
According the Maki formula~\cite{Maki1964}, $\alpha$ can be expressed as
\begin{equation}
\alpha=\frac{\sqrt{2}H_{c2}^{Orb}(0)}{H_{P}(0)}, \label{eq:2}
\end{equation}
where $H_{c2}^{Orb}(0)=0.693\times H'\times T_c$ is the upper critical field in the absence of the spin term, and $H_{P}(0)$ is the Pauli limited field. In the weakly coupled BCS scenario, we have $H_{P}^{BCS}(0)=1.84\times T_c$.
Therefore, a very simple relation $\alpha=0.53H'$ can be obtained if $H'$ is expressed in units of T/K. This relation, which is represented by the red dashed line in Fig. 4(b),
reveals the close relation between $\alpha$ and slope of $H_{c2}$ vs $T$ at $T_c$. The experimental data roughly follows the
tendency of the Maki formula. The departure of the data from the red dashed line is a consequence of the enhancement of $H_{P}(0)$ over $H_{P}^{BCS}(0)$ due to the strong coupling effect.
The present 12442 system also exhibits this tendency to a greater extent than that in the other FeSC systems.

The value of $\alpha$ can be determined using the physical quantities in the normal state~{\cite{WHH-2}:
\begin{equation}
\alpha\propto \gamma_n\rho_n, \label{eq:3}
\end{equation}
where $\gamma_n$ and $\rho_n$ are the normal state electronic specific heat (SH) coefficient and normal state dc resistivity, respectively. This may provide clues for exploring the origin of the large $\alpha$ in KCa$_2$Fe$_4$As$_4$F$_2$.
The SH measurements revealed a rather high SH jump $\Delta C/T|_{T_c}=32.5-37.5$ mJ/mol$\cdot$Fe K$^2$ in KCa$_2$Fe$_4$As$_4$F$_2$~\cite{12442-1,TWangSH}; this value is the second-largest after the value for the K-doped 122 system
Ba$_{0.6}$K$_{0.4}$Fe$_2$As$_2$ ($\sim$49 mJ/mol$\cdot$Fe K$^2$)~\cite{Mu2009} in the family of FeSCs. This suggests a large magnitude of $\gamma_n$ in KCa$_2$Fe$_4$As$_4$F$_2$, assuming a proportional relationship between $\gamma_n$
and $\Delta C/T|_{T_c}$. From the viewpoint of the normal state dc resistivity, $\rho_n$ of KCa$_2$Fe$_4$As$_4$F$_2$ is thrice (5 times) that of Ba$_{0.6}$K$_{0.4}$Fe$_2$As$_2$ at 300 K (40K)~\cite{TWang2019,Mu2009}.
In other FeSCs systems , e.g., the 1111 system, the value of $\rho_n$ is large, whereas $\gamma_n$ is small~\cite{Ma2016,ChuSH}.
Considering the above two factors synthetically and comprehensively, we find that the large value of $\alpha$ in KCa$_2$Fe$_4$As$_4$F$_2$ can be attributed qualitatively to the fact that both $\gamma_n$ and $\rho_n$ are relatively larger.

\section*{4 Conclusions}
In summary, the high magnetic field dependence of the resistivity for the KCa$_2$Fe$_4$As$_4$F$_2$ single crystals was
measured. The extremely steep increase of in-plane $H_{c2}^{ab}$ with cooling near $T_c$ changes to a bending behavior, resulting in a rapid decrease in the anisotropy $\Gamma$ with cooling.
The evolution feature of the $H_{c2}^{ab}-T$ curve can only be described by the WHH model with a large Maki parameter $\alpha=9$, indicating a strong Pauli paramagnetic effect in the present system.
The relatively large $\gamma_n$ and $\rho_n$ may explain the observed large value of $\alpha$.


\emph{This work is
supported by the Youth Innovation Promotion Association of the
Chinese Academy of Sciences (No. 2015187), the National
Natural Science Foundation of China (Nos. 11204338, 11704385 and 11874359), and
the "Strategic Priority Research Program (B)" of the Chinese
Academy of Sciences (No. XDB04040300). We thank Dr. X. Z. Xing for the help when handling the WHH model.}





\end{document}